\documentclass[a4paper,11pt]{elsarticle}
\usepackage{graphicx}
\usepackage{amssymb}
\usepackage{amsmath}
\usepackage{caption}
\usepackage{subcaption}
\usepackage{adjustbox}
\usepackage{float}
\usepackage{natbib}
\usepackage{hyperref}
\begin{document}
\begin{frontmatter}
\title{Characterizing the Effects of Randomness in the Tent Map}
\author{Dhrubajyoti Biswas \footnote{Email Address: dhrubajyoti98@protonmail.com}}
\address{Department of Physics, Indian Institute of Technology Madras, Chennai}
\author{Soumyajit Seth \footnote{Corresponding Author} \footnote{Email Address: ss14rs057@iiserkol.ac.in}}
\address{Department of Physical Sciences, Indian Institute of Science Education and Research, Kolkata}
\date{\today}

\begin{abstract}
When the parameter of a map is chosen, at each iteration step, following a certain rule, is called \textit{Parametric Perturbation}. If the parameters are drawn from a distribution, then this perturbation is called \textit{Random Parametric Perturbation}. Studies have already been done on both Periodic and Random perturbations of a continuous map.  Here, we have applied this technique on a tent map, which is a piecewise continuous map, and obtained numerical results.
\end{abstract}


\end{frontmatter}

\section{Introduction}

In the study of discrete dynamical systems, the tent map \cite{strogatz2018}\cite{crampin1994chaotic} is a well known candidate for a map which shows chaotic orbits and other typical dynamical behavior. Mathematically, the tent map, with $r\epsilon[0,2]$, is described by the recursion relation:

\begin{equation}
x_{\rm n+1}=f(x_{\rm n},r)=\begin{cases}rx_{\rm n},~0<x_{\rm n}\leq0.5\\r(1-x_{\rm n}),~0.5<x_{\rm n}\leq1
\end{cases}
\label{eqn1}
\end{equation}
 
To find the fixed points, we solve the following equation, which follows from the definition of a fixed point \cite{feldman2012}:

\begin{equation}
f(x^{\rm *},r)=x^{\rm *}
\label{eqn2}
\end{equation}

where $x^{\rm *}$ is a fixed point.
It follows from Eqn.\ref{eqn2} that that there exists one solution to this equation for $r\epsilon[0,1]$ given by $x^{\rm *}=0$, and this is an attractive fixed point. When $r\epsilon[1,2]$, there is another solution (other than $x^{*}=0$), given by: \begin{equation}x^{\rm *}=\frac{r}{1+r}
\label{eqn3}
\end{equation} In this case, both the fixed points are unstable.

\begin{figure}[tbh]
\centering
\includegraphics[width=0.7\textwidth]{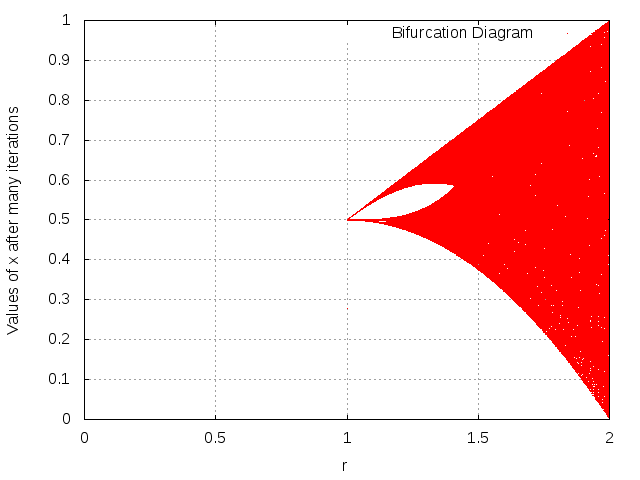}
\caption{Bifurcation Diagram of Tent Map. Note that for $r\epsilon[0,1]$, the time series dies down to zero as $n\rightarrow\infty$, whereas for $r\epsilon[1,2]$, the map becomes chaotic.}
\label{fig1}
\end{figure}

The bifurcation diagram of the tent map in the given parameter range is shown in Fig.\ref{fig1}.

All such maps, including the tent map, are deterministic systems, which means that, provided an initial starting point, we can use the recursion rule, to find out the subsequent values in the sequence. The tent map is used in many real world applications, like image encryption \cite{radwan2013image}, random number generation \cite{nejati2012discrete}\cite{lv2015perturbation}, and many other things.

In real-life, any dynamical phenomena, is `random' or `stochastic' in nature, to some extent, however small. A real-life example would be the prices of a stock in a stock market. If the price of a certain stock evolves in a deterministic way following a map, and if the starting price of the stock is constant, then the evolution of price is already known. But, it doesn't happen in the real world- even if the starting price of a stock is same, the final price of the stock after a day of trading varies widely day to day. Another such example is the case of population dynamics. Such cases justify the studies of the well known discrete dynamical maps in the 'stochastic' regime. 

Perturbation techniques have been applied to discrete maps before \cite{xing2011new} \cite{baptista1996dynamics} \cite{saratchandran1996dynamics} \cite{harikrishnan1987}. Other than adding a certain amount of noise to the value of $x$ (i.e. the state variable) at each iteration (which is called state-variable perturbation, demonstrated in \cite{baptista1996dynamics}), another way to introduce randomness into the dynamical systems governed by maps is to choose the parameter at each iteration step following some rule (the tent map is defined using a single parameter $r$, but there are numerous examples of maps being defined by using two or more parameters, the standard examples include the Henon Map \cite{henon1976two} and the Bogdanov map \cite{arrowsmith1993bogdanov}). This technique is called Parametric Pertubation, and has been applied in case of the logistic map in \cite{saratchandran1996dynamics} in the form of a periodic perturbation. Other literature exists in the form of studies of modulated discrete maps where the parameter of one map is the state variable of the other map in \cite{harikrishnan1987}. Yet another form of parametric perturbation exists through delayed feedback mechanism, which has been studied in \cite{nandi2005phase}. All these perturbations are deterministic in nature. On the other hand, Random Parametric Perturbations, where the parameters are drawn from distributions at each iteration step, incorporates stochasticity, and have been applied to the logistic map in \cite{khaleque2015effect}. In this paper, we deal with the case of Random Parametric Perturbation in case of the tent map which, unlike the logistic map which is fully continous, has a piecewise continous form.

In our case of the tent map, the parameter $r$ is chosen from the range $[q_{\rm 1},q_{\rm 2}]$, where $q_{\rm 1}\geq0$ and $q_{\rm 2}\leq2$, with the condition that $q_{\rm 1}\leq q_{\rm 2}$. The distributions from which the parameter has been sampled are the uniform distribution and the symmetric triangular distribution. We have seen from the bifurcation diagram of the tent map that when the value of $r$ is less than one, the time-series dies down to zero as because zero is an attracting fixed point in that parameter range, which prompts us to set even a further limit on $q_{\rm 1}$, which is $q_{\rm 1}\geq1$. Note that, this is the chaotic region for the tent map. 

One has to be careful with the choice of the distribution which is used to draw the parameters from, such that the distribution terminates at $q_{\rm 1}$ and $q_{\rm 2}$. On a deeper level, the choice of the distribution is determined by the real world problem which we wish to model. 

\section{How do we quantify the dynamics?}
The random tent map, just like it's non-random counterpart, maps the interval $[0,1]$ onto itself, given $r\epsilon[0,2]$. In the random tent-map, the values $x$ can take lies between a certain $x_{\rm min}$ and $x_{\rm max}$, and we can define a quantity $\Delta x=x_{\rm max}-x_{\rm min}$, which should depend on the values of $q_{\rm 1}$ and $q_{\rm 2}$. This is characteristically different from the case of a non-random map, where the values of $x$ tends to a certain number or oscillates about/move away depending on the type of fixed point we are considering, whereas for the random map, $x$ is ergodic between the aforementioned $x_{\rm min}$ and $x_{\rm max}$. This has been demonstrated in Section 3.2.

We can also define another quantity $\Delta_{\rm n}=|x_{\rm 1}(n)-x_{\rm 2}(n)|$, where $x_{\rm 1}(n)$ and $x_{\rm 2}(n)$ are the sequences generated from the same tent-map using the same set of parameter values $r$, but with slightly different initial values. This quantity can also be calculated by another approach, where we begin with the same initial conditions, but the parameter for each evolution is different, but drawn from the same distribution. This method is called the `Nature vs Nurture' (NVN) method and the previous method is known as the `Traditional Method' (TM) in existing literature\citep{RePEc}. 

Mathematically, in the TM method, we take two initial values for $x$, such as $x_{\rm 0}$ and $x_{\rm 0}+\delta$, where $\delta$ is small. Then, the map $f(x,r)$ is applied to both of them iteratively to generate two different sequences, where $r$ is chosen from a certain distribution at each iteration. In contrast, in the NVN method, a single initial condition is chosen for $x$, say $x_{\rm 0}$. Now, two maps are applied to it, $f(x,r_{\rm 1})$ and $f(x,r_{\rm 2})$ to generate two different sequences, where $r_{\rm 1}$ and $r_{\rm 2}$ are different numbers sampled from the same distribution.

It is necessary that for calculation of values of such quantities, an average over many individual evolutions or pair of evolutions as may be needed (for the TM and NVN methods) to smooth out any statistical fluctuations. For our calculations, we have set the total number of evolutions/pair of evolutions carried out before taking the average to 10000.
\section{Numerical Results}
From eqn.\ref{eqn3} , we can approximately guess that, for the random tent map:
\begin{equation}
<x_{\rm n\rightarrow\infty}>=\frac{\overline{r}}{1+\overline{r}}
\label{eqn4}
\end{equation}where $\overline{r}$ is the mean of the distribution used to draw the parameter values.
\subsection{General Results}
\begin{figure}[tbh]
\centering
\begin{subfigure}[b]{0.5\textwidth}
\includegraphics[width=1.0\textwidth]{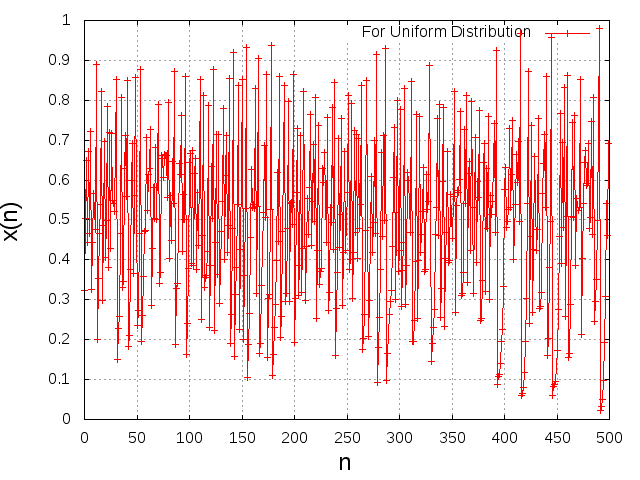}
\caption{Time Series using Uniform Distribution}
\end{subfigure}%
\begin{subfigure}[b]{0.5\textwidth}
\includegraphics[width=1.0\textwidth]{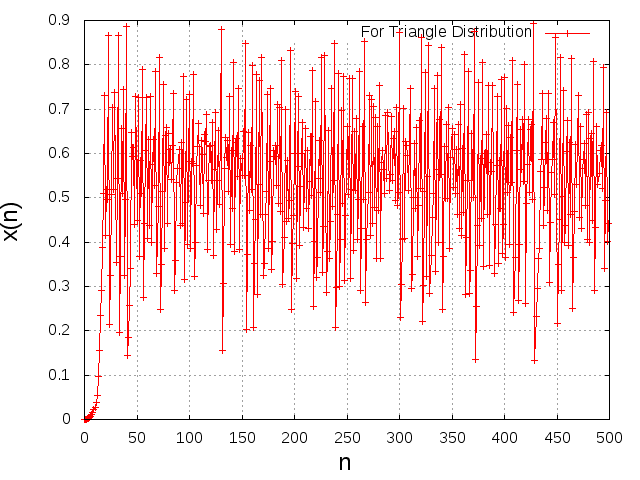}
\caption{Time Series using Triangular Distribution}
\end{subfigure}
\caption{Time Series generated using Uniform and Triangular Distribution. Note that, in this particular figure, $n$ has been substituted with $t$, which denotes time/iteration number.}
\label{fig2}
\end{figure}

In Fig.\ref{fig2}, the time series plots generated using both the distributions is shown.
Using the ansatz of eqn.\ref{eqn4} (the second column of Table.\ref{table1}), and along with actual numerical calculations (the last column of Table.\ref{table1}), the average values of $x$ is calculated and they are tabulated in Table.\ref{table1} . The deviations of the actual values obtained numerically from the predicted values are smaller in the case of the Triangular distribution compared to the Uniform distribution. A similar result was obtained for the logistic map in \cite{khaleque2015effect}.

\begin{table}
\centering
\begin{tabular}{|c|c|c|}
\hline
Distribution,  $\mathbf r\epsilon[q_{1},q_{2}]$ & Theoretical & Numerical \\
\hline
Uniform, $[1,2]$ & 0.6 & 0.499 \\
\hline
Uniform, $[1.5,2]$ & 0.636 & 0.522 \\
\hline
Sym. triangular,$[1,2]$ & 0.6 & 0.540 \\
\hline
Sym. triangular, $[1.5,2]$ & 0.636 & 0.566 \\
\hline
\end{tabular}
\caption{Steady State Values of $x$}
\label{table1}
\end{table}

This can be attributed to the fact that for the case of the uniform distribution, all values of the parameter in the range $[q_{\rm 1},q_{\rm 2}]$ is equally probable resulting in a higher deviation from the expected values, whereas in the case of the symmetric triangular distribution, the values near the peak of the distribution (which also denotes the average because the distribution is symmetric about this point, i.e. near $r=\overline{r}$) are more likely to be drawn than the values lying further away on both sides of the peak.
\subsection{Ergodicity of the Random Map and The “Min-Max” Diagrams} 
\begin{figure}[h]
\centering
\begin{subfigure}[b]{0.5\textwidth}
\includegraphics[width=1.0\textwidth]{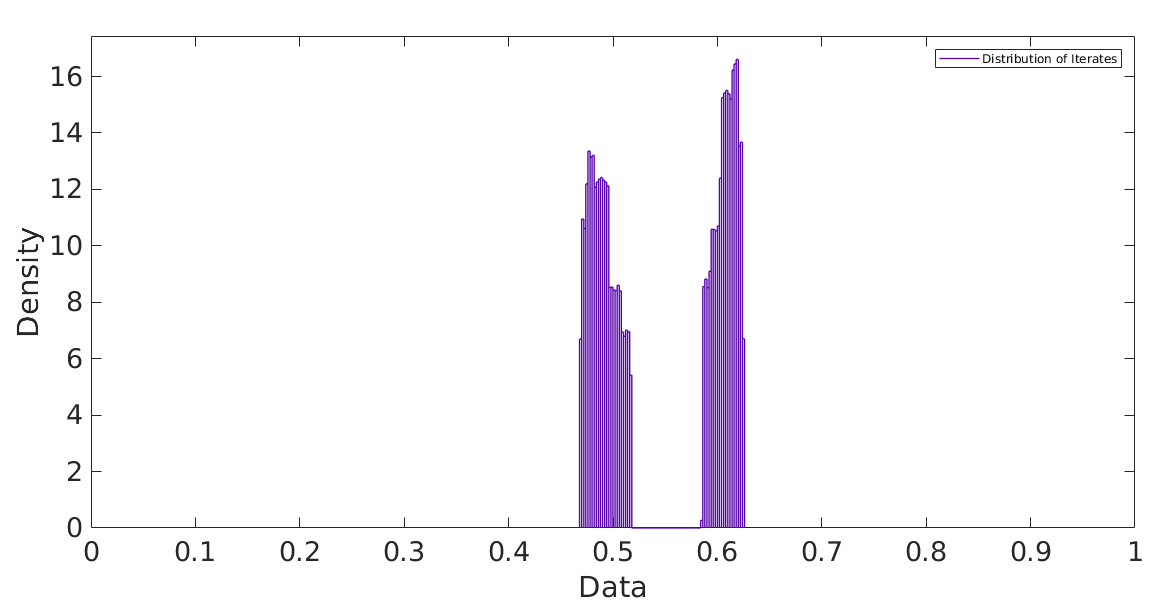}
\caption{Distribution for the Non-Random Tent Map with $r=1.25$}
\label{fig3a}
\end{subfigure}%
\begin{subfigure}[b]{0.5\textwidth}
\includegraphics[width=1.0\textwidth]{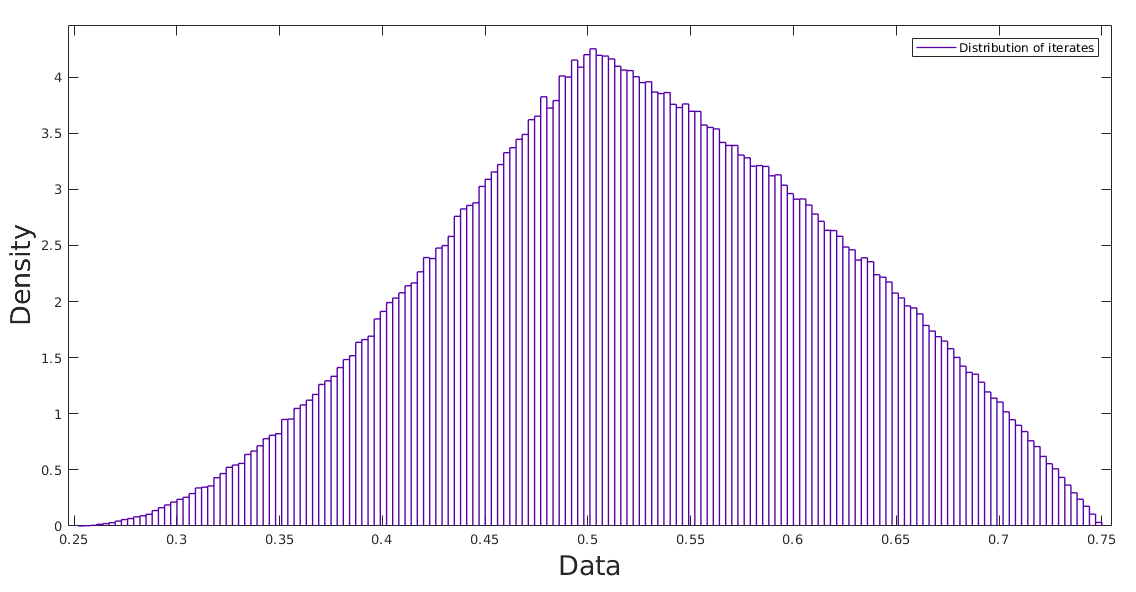}
\caption{Distribution for the Random Tent Map with $q_{\rm 1}=1.0$ and $q_{\rm 2}=1.5$}
\label{fig3b}
\end{subfigure}
\caption{Distributions of iterates for Random and Non-Random Maps. Note, how in the case of the non-random map, the values of the iterates are localized around some values, whereas in the case of the random map, the iterates are distributed between a certain minimum and maximum value, which is determined by the range $[q_{\rm 1},q_{\rm 2}]$ of the distribution.}
\end{figure}

\begin{figure}[tbh]
\begin{subfigure}[b]{0.5\textwidth}
\centering
\includegraphics[width=1.0\textwidth]{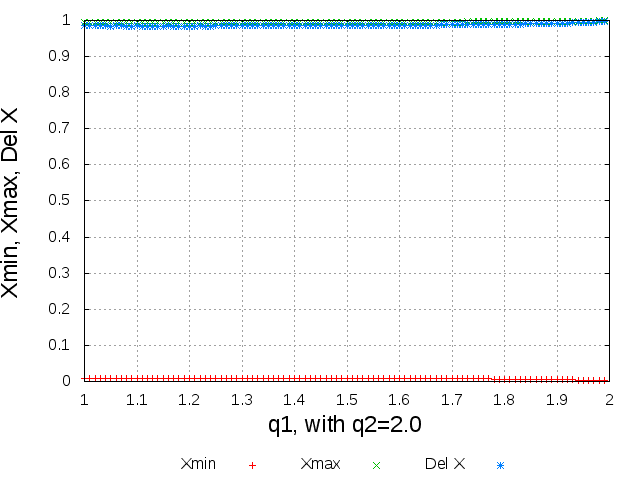}
\caption{Min-Max diagram for Uniform Distribution, by varying $q_{\rm 1}$}
\label{fig4a}
\end{subfigure}%
\begin{subfigure}[b]{0.5\textwidth}
\centering
\includegraphics[width=1.0\textwidth]{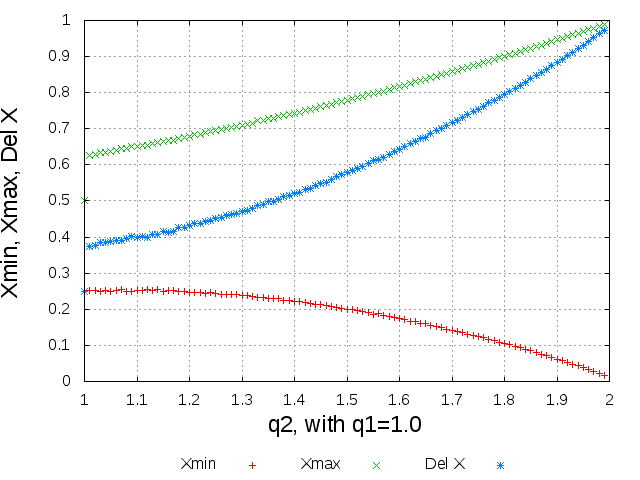}
\caption{Min-Max diagram for Uniform Distribution, by varying $q_{2}$}
\label{fig4b}
\end{subfigure}
\caption{The Min-Max Diagrams for the Uniform Distribution}
\end{figure}

Ergodicity of the value of $x$, mentioned in Section 2, has been shown here. Firstly, the distribution of the iterates has been plotted for the non-random tent map with $r=1.25$ in Fig.\ref{fig3a} whereas in Fig.\ref{fig3b} we have plotted the distribution of the iterates for the random map, using the flat distribution for $q_{1}=1.0$ and $q_{2}=1.5$. This was the preferred choice because the mean of the uniform distribution is $\overline{r}=1.25$, which would coincide with the $r=1.25$ case of the non-random map.
We see that, for the non-random map, the distribution of the iterates are localized about two points on either side of the fixed point, whereas for the random map, we see that the iterates are ergodic between two values of $x$, given by $x=x_{\rm min}$ and $x=x_{\rm max}$. In our particular example, we have chosen the value of $r=1.25$ for the non-random map, which gives rise to a fixed point at $x^{\rm *}=0.56$, whereas, for the random map, we have chosen $q_{\rm 1}=1.0$ and $q_{\rm 2}=1.5$ with a flat distribution, such that $\overline{r}=1.25$. We see, for the random map, $x_{\rm min}\backsimeq0.2$ and $x_{\rm max}\backsimeq0.75$, with a peak near $x=0.5$, which is consistent and has been further verified by the Min-Max diagrams as shown below.
We can plot the values of $\Delta x$, $x_{\rm min}$ and $x_{\rm max}$ in two different ways: we can set $q_{\rm 2}=2.0$ and vary $q_{\rm 1}$ in the range $[1,2]$ (Fig.\ref{fig4a}) and similarly, set $q_{\rm 1}=2.0$ and vary $q_{\rm 1}$ in the range $[1,2]$ (Fig.\ref{fig4b}). The plots (containing the variation of $\Delta x$, $x_{\rm min}$ and $x_{\rm max}$, or the “Min-Max" Diagrams) have been shown for the uniform distribution only and the plots for the triangular distribution are approximately same. From this, we can infer that the Min-Max diagrams obtained is the property of the `random' map itself and not related to the use of any specific distribution for drawing the parameters. Thus, the Min-Max diagrams can be used to characterize the random maps.

\subsection{Variation of $\Delta_{n}$ in the TM Method}
\begin{figure}[tbh]
\begin{subfigure}[b]{0.5\textwidth}
\centering
\includegraphics[width=1.0\textwidth]{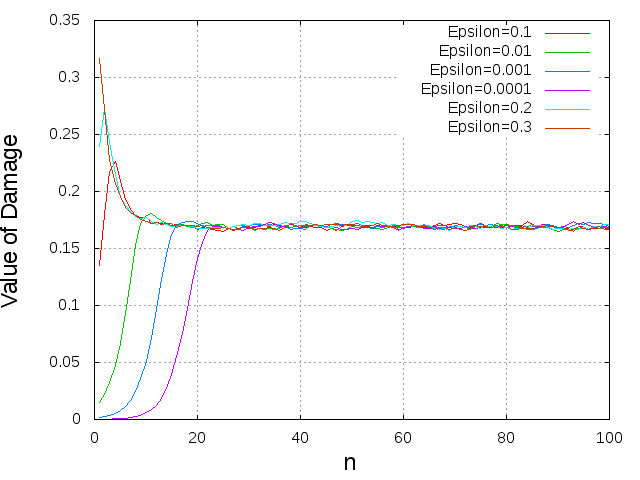}
\caption{$\Delta_{\rm n}$ vs $n$ for Uniform Distribution}
\label{fig5a}
\end{subfigure}%
\begin{subfigure}[b]{0.5\textwidth}
\centering
\includegraphics[width=1.0\textwidth]{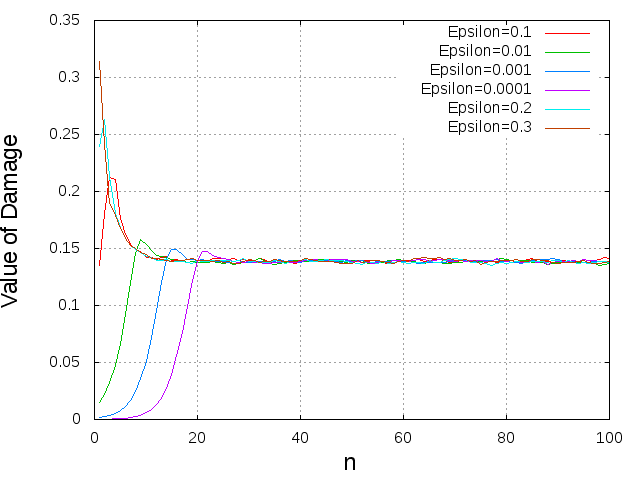}
\caption{$\Delta_{\rm n}$ vs $n$ for Triangular Distribution}
\label{fig5b}
\end{subfigure}
\caption{Evolution of $\Delta_{\rm n}$ (`Damage') as a function of $n$ (TM Method)}
\end{figure}

\begin{figure}[H]
\begin{subfigure}[b]{0.5\textwidth}
\centering
\includegraphics[width=1.0\textwidth]{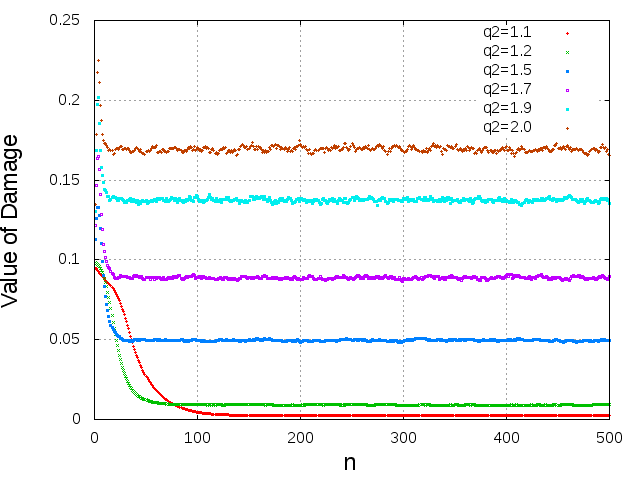}
\caption{For the Uniform Distribution}
\label{fig6a}
\end{subfigure}%
\begin{subfigure}[b]{0.5\textwidth}
\centering
\includegraphics[width=1.0\textwidth]{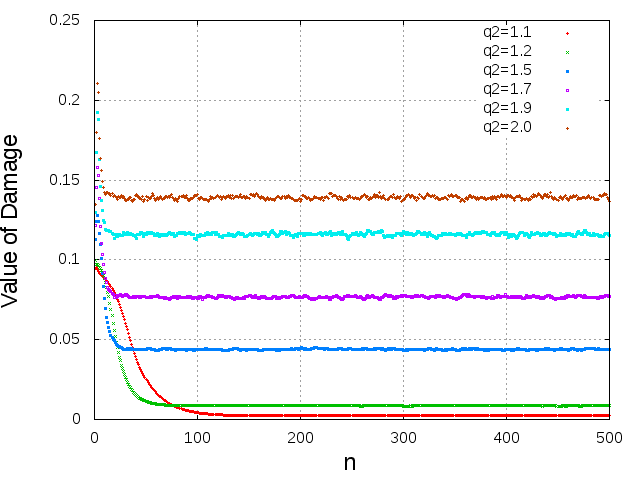}
\caption{For the Triangular Distribution}
\label{fig6b}
\end{subfigure}
\caption{Plots showing saturation value of damage $\Delta_{\rm n(sat)}$ for $q_{\rm 1}=1.0$ and $q_{\rm 2}\epsilon[1,2]$ (TM Method)}
\end{figure}

Using the TM method described in Section 2, the value of the variable $\Delta_{\rm n}$, also called `Damage' (described in\cite{RePEc}), can be plotted as a function of $n$, i.e. the number of steps taken in the iteration. We take the initial separations to be $0.1$, $0.01$, $0.001$, $0.0001$, $0.2$ and $0.3$ (denoted as `Epsilon' on the plots). Fig.\ref{fig5a} shows the evolution of $\Delta_{\rm n}$ for the Uniform Distribution of $r$ while Fig.\ref{fig5b} shows the evolution for the Triangular Distribution. The range $[q_{\rm 1},q_{\rm 2}]$ has been set to $[1,2]$.
It is seen that, apart from small fluctuations, the value of $\Delta_{\rm n}$ goes to a saturation value (some $\Delta_{\rm n(sat)}$) as $n\rightarrow\infty$. The saturation value obtained, i.e. $\Delta_{\rm n(sat)}$ depends on the values of $q_{\rm 1}$ and $q_{\rm 2}$. This can be shown by keeping $q_{\rm 1}=1.0$ and then varying $q_{\rm 2}$ in the range $[1,2]$. This has been shown for both Uniform (Fig.\ref{fig6a}) and Triangular (Fig.\ref{fig6b}) distributions.

\subsection{Variation of $\Delta_{\rm n}$ in the NVN Method}
\begin{figure}[H]
\begin{subfigure}[b]{0.5\textwidth}
\centering
\includegraphics[width=1.0\textwidth]{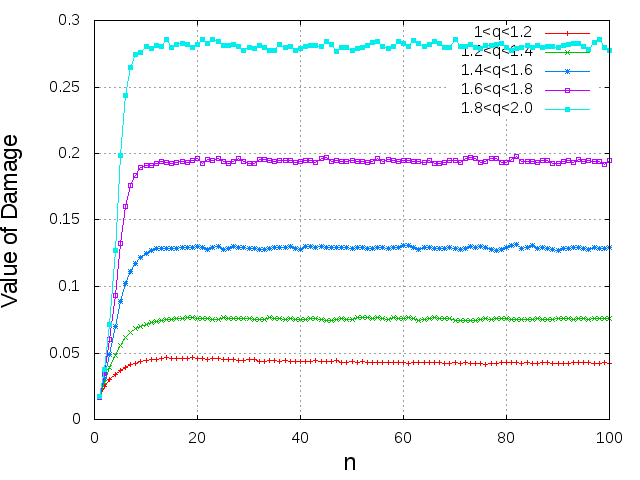}
\caption{$\Delta_{\rm n}$ vs $n$ for the uniform distribution}
\label{fig7a}
\end{subfigure}%
\begin{subfigure}[b]{0.5\textwidth}
\centering
\includegraphics[width=1.0\textwidth]{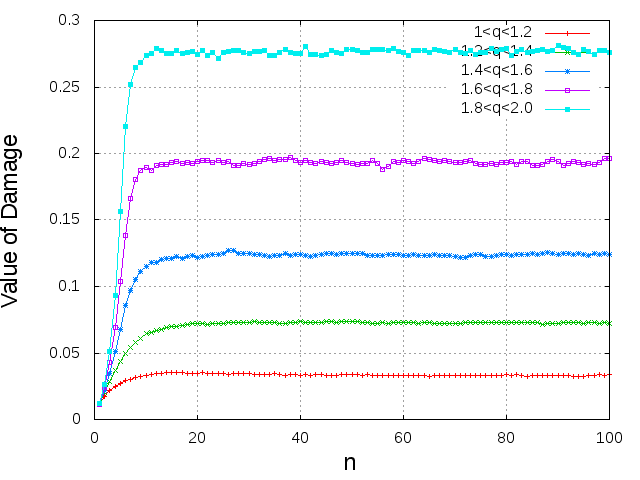}
\caption{$\Delta_{\rm n}$ vs $n$ for the triangular distribution}
\label{fig7b}
\end{subfigure}
\caption{Evolution of $\Delta_{\rm n}$ (`Damage') as a function of $n$ (NVN Method)}
\end{figure}

For the NVN method, we plotted $\Delta_{\rm n}$ as function of $n$. It is seen that, the behaviors remain the same, in the sense that at large values of $n$, the variable $\Delta_{\rm n}$ reaches a saturation value, similar to that of the TM method, but the final saturation value reached is different than what was obtained in the TM case. Fig.\ref{fig7a} shows the variation for the Uniform distribution and Fig.\ref{fig7b} shows the variation for the Triangular Distribution. Both the plots were generated for fixed sub intervals of $q_{\rm 1}$ and $q_{\rm 2}$, as denoted on the plots. 

\section{Conclusion}
In this paper, the characterization of the dynamics of the chaotic tent-map was attempted in the `stochastic' regime. It is observed that, as shown in the case for the logistic map in \cite{khaleque2015effect}, the tent map also shows ergodic behaviour in $x$ while in the stochastic regime. The range of ergodicity of $x$ is limited by a certain $x_{\rm min}$ and $x_{\rm max}$, which strongly depends on $q_{\rm 1}$ and $q_{\rm 2}$. This has been established using the ``Min-Max" diagrams, which are seen to remain invariant with respect to the type of distribution used, which makes it a property of the map itself in the `stochastic' regime, and thus can be used to characterize the dynamics obtained independently.

The long term behaviour of $x$ also depends on the type of distribution used and agreements with mean values are more in case of the triangular distribution than that of the uniform distribution. The value of `damage' or the difference between the values of iterations obtained at same time-step but from two different methods (TM and NVN) are seen to approach a constant as $n\rightarrow\infty$, but the saturation value obtained depends on the range of the distribution used, and the method applied to calculate the damage (i.e. TM or NVN methods), but is independent of the initial conditions imposed.

Open questions remain, on how to arrive at analytic results which would support our numerical calculations. Another interesting question would be regarding the choice of the distribution used. In this study (and also in \citep{khaleque2015effect}), the distributions were arbitrarily chosen, keeping in mind only the range of the parameter to be drawn. But in the real world, to actually find the distribution which would actually model real-life data is something which isn't clearly understood and requires further investigation.

\section*{Acknowledgements}
The work of DB was supported by IISER Kolkata Summer Research Fellowship SRP18-0166. The work of SS was supported by DST-INSPIRE, Government of India (Ref. No: IF150667).

\section*{References}
\bibliography{paper_report}

\end{document}